\documentclass[nofootinbib,showpacs,preprintnumbers,superscriptaddress] {revtex4}
\usepackage{amssymb,amsfonts,amsmath,graphicx,}
\usepackage{color}
\usepackage{enumerate}
\usepackage{braket}
\usepackage{epstopdf}

\usepackage{varioref}
\usepackage{slashed}

\begin{document}

\title{Bosonic dark matter in a coherent state driven by thermal fermions}

\author{Eung Jin Chun}
\email{ejchun@kias.re.kr}
\affiliation{Korea Institute for Advanced Study, Seoul 02455, Korea}

\begin{abstract}
If there is a scalar boson field interacting dominantly with a quark or a lepton in the thermal background,
its coherent oscillation can be generated through the thermal effect and becomes a good dark matter candidate in a wide range of the coupling and masses. We describe the general features of this mechanism and analyze how it works in various situations considering asymptotic limits where analytic solutions can be obtained.
\end{abstract}

\maketitle

\section{Introduction}

One of attractive dark matter candidate is a light boson in a coherent state.
A classic example is the QCD axion whose
cosmological population arises from the misalignment mechanism which drives the coherent oscillation of the axion field around the minimum of the effective QCD potential \cite{aDM}.
Another interesting example is a ultra-light bosonic condensate whose de Broglie wavelength extends to a galactic scale affecting the dark matter halo and structure formation \cite{fDM}.  Its abundance
could also be due to the misalignment production by a certain effective potential generated, e.g., by a gluon condensate \cite{cchan}.

In this article, we discuss a novel mechanism for the generation of a coherent bosonic state in a thermal background \cite{esteban,batell}. Suppose that a scalar boson couples to a thermal fermion (quark or lepton). Then its thermal distribution generates temperature-dependent mass  and source term in the classical equation of motion governing the cosmological evolution of the coherent state. As a result, it may constitutes  the observed dark matter density depending on parameters of the  model.  Some earlier works on thermal effect in scalar field dynamics can be found in \cite{dine,anisimov,ramazanov}.
In Section II, we describe the general features of the cosmological evolution of a classical boson field coupling to thermal fermions in connection with the freeze-in mechanism and Big Bang Nucleosynthesis (BBN) constraints.  In Sections III and IV, approximate analytic solutions are derived considering various limits depending on the sizes of the Yukawa coupling and masses of the scalar boson field and the thermal fermion.
Finally, we conclude in Section V.

\section{Thermal effect in the evolution of a boson field}

We assume a light boson $\phi$ with mass $m_\phi$ that couples to a Standard Model (SM) fermion through a tiny Yukawa coupling $y_\phi$:
\begin{equation} \label{phi-f-f}
{\cal L}_{\phi f f} =  y_\phi \phi (\bar f_L f_R + \bar f_R f_L)
\end{equation}
where $f$ could be a quark $q$, charged-lepton $l$, or neutrino $\nu$.
This interaction is supposed to arise after the electroweak symmetry breaking from a higher dimensional operator  $\phi {\cal L}_{\rm SM}/\Lambda$ where $\Lambda$ is a cutoff scale and ${\cal L}_{\rm SM}=\bar f_L \Phi f_R + h.c.$ is a SM Higgs-fermion-fermion operator.
In the thermal background of the fermions, the classical boson field $\phi(t)$ describing the homogeneous coherent oscillation follows the evolution equation:
\begin{equation}
\ddot{\phi}(t)+3H\, \dot\phi(t) + (m_\phi^2 +m_T^2)\, \phi(t) = {\partial \over \partial \phi} \langle{\cal L}_{\phi f f} \rangle_T
\end{equation}
where $H$ is the Hubble parameter, and the thermal effect encoded in  $m_T^2$  and  $\langle{\cal L}_{\phi f f} \rangle_T$ arises from the thermal distribution of the fermion $f$.
Following the calculation in a finite temperature \cite{dolan}, one obtains
\begin{equation}
m_T^2 =   \frac{ g_f }{24} y_\phi^2 T^2 , ~~\mbox{and}~~
\langle{\cal L}_{\phi f f} \rangle_T= y_\phi \phi  \left[ {g_f m_f T^2 \over 24} \right]
\end{equation}
where $g_f$ is the degrees of freedom of $f$: $g_f=4 N_c$ with the color factor $N_c=3\, (1)$ for quarks  (charged leptons) $f=q \,(l)$, and $g_f=2$ for  Majorana neutrinos $f=\nu$.  The above equation is valid in the regime of $T\gg |m_f-y_\phi \phi|$ where the thermal fermion remains relativistic. Indeed, this condition is satisfied by all the solutions found in the next sections.

\medskip

In the era of radiation domination, we have the temperature-time relation $T^2=c_t^2 M_P/2t$ with the reduced Planck mass $M_P \approx 2.4\times 10^{18}$ GeV and $c_t \approx 1.74/g_*^{1/4}$. Thus the evolution equation of the classical field $\tilde \phi \equiv \phi/M_P$ can be written in terms of $x\equiv m_\phi t$ as follows:
\begin{equation} \label{eq:phi}
\begin{split}
&\tilde \phi''(x) +  {3 \over 2 x}\tilde\phi'(x) + \left(1+ {x_1\over x}\right) \tilde\phi(x)  =  {x_S\over x} \\
 \mbox{where}~~&  x_1 \equiv y_\phi^2 {c_t^2 g_f \over 48} {M_P\over m_\phi}~~\mbox{and} ~~
x_S \equiv y_\phi {c_t^2 g_f \over 48} {m_f\over m_\phi}.
\end{split}
\end{equation}
Note that this equation is valid after the electro-weak phase transition at $T_{ew} \approx 100$ GeV until the fermion $f$ remains relativistic in thermal equilibrium. That is, the thermal effect is switched off below $T \approx T_f = m_f$, around which the charged fermions $f=q,l$ become non-relativistic and annihilate away, e.g., through $f\bar f\to \gamma\gamma$.
That is, $x_{1,S}\to 0$ for $x> x_f \equiv m_\phi t_f=c_f^2 m_\phi M_p/2m_f^2$. For simplicity, we will take a sudden change approximation that we apply Eq.~(4) from $x_{ew}$ to $x_f$, and take $x_{1,S}=0$ for $x> x_f$.  For neutrinos ($f=\nu$), which are light and in thermal equilibrium until very late,
Eq.~(4) is valid from $x_{ew}$ down to the matter-radiation equality $x_{eq}\equiv m_\phi t_{eq}$
corresponding to the temperature $T_{eq} \approx 0.81$ eV.

It is useful to note
\begin{equation}
\begin{split}
&x_{ew}\approx 1\times 10^{-15} c_{ew}^2 m_{20}, ~~~
x_{eq}\approx 1.8 \times 10^7 c_{eq}^2 m_{20}; \\
& x_1=  5.1\times 10^{46} g_f y_\phi^2 c_1^2  m_{20}^{-1},~~~
 x_f =4.9 \times 10^{-4} c_f^2  m_{20} \left( m_e \over m_f \right)^2
\end{split}
\end{equation}
where $m_{20}\equiv m_\phi/10^{-20}\,\mbox{eV}$.
To form the coherent state of the bosonic dark matter well before the matter-radiation equality, we require $x_1, x_f \ll x_{eq}$.
Depending on the hierarchy among $x_1, x_f$ and $x_{ew}$, we will have different type of solutions.
In each case, the bosonic energy density in the form of coherent oscillation at $x\to \infty$ can be expressed as
\begin{equation} \label{eq:rhof}
\rho_\phi(x) \approx {c_f^2 x_S^2 \over \pi} {m_\phi^2 M_P^2 \over x^{3/2}}
\end{equation}
where $c_f$ is a function of $x_1$ and $x_f$ specific to the  thermal fermion $f$.
If it constitutes the whole dark matter density, we need  the relation: $\rho_{\phi}(x_{eq}) \approx 0.48 \mbox{ eV}^4$ leading to
\begin{equation} \label{eq:cf}
 c_f^2 x_S^2 \approx 2.5 \times 10^{-4} m_{20}^{-1/2} .
 \end{equation}
From this, we will identify the parameter regions consistent with the observed dark matter density in various limiting cases for given quarks and leptons.  Throughout our analysis, we will neglect the mild variation of $c_t$ which is taken to be 1.

As we will see, the right abundance of the bosonic dark matter typically requires  tiny couplings for which
the freeze-in mechanism  \cite{hall} can be operative.  The dominant channel for the freeze-in production is $f\bar f \to \phi \gamma$ or $\phi g$ involving the electromagnetic or QCD  coupling. The resulting number density normalized by the entropy density is given roughly by
$
Y_\phi  \sim  {\alpha_y \alpha_f M_P/ T}
$
evaluated at $T=m_f$ with $\alpha_y \equiv y_\phi^2/4\pi$ and $\alpha_f = \alpha_{em}$ ($\alpha_c$) for charged leptons (quarks).  Thus, assuming negligible freeze-in abundance in our analysis, we will require
\begin{equation} \label{freezein}
y_\phi \ll { 4.5\times  10^{-11} \over \sqrt{\alpha_f} } \sqrt{ m_f \over m_\phi } .
\end{equation}
There is another restriction applicable for $m_\phi \lesssim1$ MeV.
Thermalization of a light boson could be in conflict with the standard prediction of BBN. 
To avoid it, we make a rough requirement that the interaction rate of the Yukawa coupling smaller than the Hubble expansion rate, that is, $\alpha_y \alpha_f T \lesssim T^2/M_P$ at $T=m_f$. This gives us
\begin{equation} \label{bbn}
y_\phi \lesssim {5.1\times 10^{-11} \over \sqrt{\alpha_f}}  \sqrt{m_f \over m_e}.
 \end{equation}
These conditions will remove some of our parameter space for the coherent bosonic dark matter.

\section{Coherent oscillation from charged fermions}

Depending on the values of the parameters $y_\phi, m_\phi$ and $m_f$, one can consider three different cases: (I) $x_1 < x_{ew} < x_f$; (II) $x_{ew} < x_f < x_1$; (III) $x_{ew}<x_1<x_f$. For each case, approximate analytic solutions can be found taking the sudden change approximation, that is, ensuring the continuity of two asymptotic solutions at  the boundary.

\medskip
\underline{Case I : $x_1 <x_{ew} < x_f$}
\medskip

This is the case where the condition $m_\phi^2 \gg m_T^2$ is maintained all the time, and thus two different equations are hold in the asymptotic regions as follows:
\begin{equation} \label{eqI}
\begin{split}
 \tilde \phi''(x) +  {3 \over 2 x}\tilde\phi'(x) + \tilde\phi(x)  \approx {x_S\over x}
 & ~~\mbox{for}~~ x_{ew}\ll x \ll x_f \\
 \tilde \phi''(x) +  {3 \over 2 x}\tilde\phi'(x) + \tilde\phi(x)  = 0
 & ~~\mbox{for}~~ x_f \ll x\,.
\end{split}
\end{equation}
Assuming the vanishing initial condition, $\tilde\phi=\tilde\phi'=0$ at $x_{ew}$, one obtains the two asymptotic solutions:
\begin{equation} \label{solI}
\tilde\phi(x)= \left\{
\begin{split}
& x_S \left[  {\pi \over \Gamma({3\over4})}
  {J_{1\over4}(x) \over (2 x)^{1/4}} \left( G_1(x)-G_1(x_{ew})\right)
-{4\Gamma(\frac{3}{4})\over3}
{J_{-{1\over4}}(x)\over (2x)^{1/4}}  \left(G_2(x)-G_2(x_{ew})\right)
\right]
 ~~\mbox{for}~~ x_{ew}\ll x \ll x_f \\
&C_1 {J_{1\over4}(x) \over x^{1/4}} + C_2 {Y_{1\over4}(x) \over x^{1/4}}
 ~~\mbox{for}~~ x_f\ll x
\end{split} \right.
\end{equation}
where $G_1(x)\equiv x\,  _1F_2({1\over2}; {3\over4}, {3\over2}; -{x^2\over4}) $ and $G_2(x) \equiv  x^{3\over2}\,  _1F_2({3\over4}; {5\over4}, {7\over4}; -{x^2\over4} )$. Here $J_n$ and $Y_n$ are Bessel functions, and $_1F_2$ denotes the hypergeometric function.
In the limit $x\to \infty$, the energy density of the boson field is described by
\begin{equation}
\rho_{\phi}(x) \approx { C_1^2 +C_2^2 \over \pi } {m_\phi^2 M_P^2\over x^{3/2} }
\end{equation}
which can be compared with (\ref{eq:rhof}) to get $c_f^2 = (C_1^2 + C_2^2)/x_S^2$.
Two integration constants $C_{1,2}$ obtained by matching the boundary condition at $x_f$ are
\begin{equation} \label{CsolI}
\begin{split}
& {C_1 \over x_S} = {\pi \over 2^{1\over4} \Gamma({3\over4})} G_1(x_f)
                      - {2^{5\over4} \Gamma({3\over4}) \over3} G_2(x_f) \\
 & {C_2 \over x_S} =  {2^{5\over4} \Gamma({3\over4}) \over3} G_2(x_f)
\end{split}
\end{equation}
taking $x_{ew} \to 0$.
Considering the limiting values of $x_f$, one obtains simple functional relations like
\begin{equation}
c_f^2 \approx  \left\{\begin{split}
& 4\sqrt{2}\, \Gamma\big({5\over4}\big)^2 x_f^2 &\mbox{for}~~ x_f \ll 1,  \\
& {\pi \Gamma({3\over4})^2 \over 2\sqrt{2}}  &\mbox{for}~~ x_f \gg 1 .
 \end{split}\right.
\end{equation}
Applying the relation (\ref{eq:cf}), we find that  the light boson  in the coherent state can explain the observed dark matter density in the following parameter regions:
\begin{equation} \label{ymI}
\begin{split}
&   y_\phi \approx 3.6 \times 10^{-24} {m_{20}^{-{1\over4}} \over N_c} {m_f \over m_e}
~~  \mbox{with}~~ {550 \over N_c^{2\over5}}
\left(m_f \over m_e\right)^{4\over5} \ll  m_{20}\ll
 2\times10^3 \left(m_f \over m_e\right)^2
 ~~\mbox{for}~~x_f \ll 1, \\
&  y_\phi \approx 3 \times 10^{-27} {m_{20}^{3\over4} \over N_c} {m_e \over m_f}
~~  \mbox{with}~~ m_{20} \gg \mbox{Max}\left[ {2 \times 10^{16} \over N_c^2}
\left(m_e \over m_f\right)^4 ,~ 2\times10^3 \left(m_f \over m_e\right)^2\right]
~~\mbox{for}~~x_f \gg 1.
\end{split}
\end{equation}
Note that the restrictions on $m_{20}$ are required to satisfy the conditions:
$x_1 \ll x_{ew}$, and $x_f\ll1$ or $\gg1$.
For light quarks ($f=u,d,s$), the above formulae are valid down to the QCD phase transition at $T_{QCD} \approx 150$ MeV (with $N_c=3$) when they condense to mesons which disappear shortly after. Ignoring  such a detail, we can take a rough condition, $m_f=150$ MeV, and apply the above equations.
The larger values of $m_\phi$ are allowed considering the bottom quark ($f=b$) which is the heaviest quark applicable to our scenario.

In the second case of (\ref{ymI}), larger values of $y_\phi$ and $m_\phi$ overlap with the freeze-in range.
Excluding this regime (\ref{freezein}), we put the following restrictions:
\begin{equation}
m_\phi \ll 46
 \mbox{ eV}\, (5.7 \mbox{ MeV})~~\mbox{for}~~ f=e~(b),
\end{equation}
requiring $y_\phi \ll 5.3 \,(1.4) \times 10^{-11}$.

\medskip
\underline{Case II: $x_{ew} < x_f < x_1$}
\medskip

In this case, $m_T^2 \gg m_\phi^2$ all the time until the thermal effect disappears at $x_f$.  Thus, the two asymptotic equations are given by
\begin{equation}
\begin{split}
 \tilde \phi''(x) +  {3 \over 2 x}\tilde\phi'(x) + {x_1\over x} \tilde\phi(x)  \approx {x_S\over x}
& ~~\mbox{for}~~ x_{ew} \ll x\ll x_f \\
 \tilde \phi''(x) +  {3 \over 2 x}\tilde\phi'(x) + \tilde\phi(x)  = 0
& ~~\mbox{for}~~ x_f\ll x .
 \end{split} .
\end{equation}
The corresponding solutions are
\begin{equation}
\tilde\phi(x)= \left\{
\begin{split}
& {x_S\over x_1} \left( 1-\sqrt{x_{ew}\over x} \cos[2(\sqrt{x x_1}-\sqrt{x_{ew}x_1})]-
{1\over 2 \sqrt{x x_1} } \sin[2(\sqrt{x x_1}-\sqrt{x_{ew}x_1})] \right)
 ~~\mbox{for}~~ x_{ew}\ll x \ll x_f \\
&C_1 {J_{1\over4}(x) \over x^{1/4}} + C_2 {Y_{1\over4}(x) \over x^{1/4}}
 ~~\mbox{for}~~ x_f\ll x ,
\end{split} \right.
\end{equation}
where
\begin{equation} \label{CsolI}
\begin{split}
& {C_1 \over x_S} =- {\pi x_f^{1\over4} \over 4 x_1} \left[ 2 x_f Y_{5\over4}(x_f)
     - Y_{1\over4}(x_f)\cos(2\sqrt{x_1 x_f})
     + \sqrt{ x_f\over x_1} Y_{-{3\over4}}(x_f) \sin(2\sqrt{x_1 x_f}) \right] ,
  \\
 & {C_2 \over x_S} =   {\pi x_f^{1\over4} \over 4 x_1} \left[ 2 x_f J_{5\over4}(x_f)
     - J_{1\over4}(x_f)\cos(2\sqrt{x_1 x_f})
     + \sqrt{ x_f\over x_1} J_{-{3\over4}}(x_f) \sin(2\sqrt{x_1 x_f}) \right] .
\end{split}
\end{equation}
Thus, one finds the simple expressions in the two limits as follows:
\begin{equation}
c_f^2 \approx \left\{
\begin{split}
&4\sqrt{2}\, \Gamma\big({5\over4}\big)^2 x_f^2   ~~\mbox{for}~~  x_f \ll x_1 \ll 1, \\
&{\pi x_f^3 \over 2x_1^2 }   ~~~\mbox{for}~~ 1\ll  x_f \ll x_1.
\end{split}\right.
\end{equation}
From this, we obtain
\begin{equation} \label{y-m-II}
\begin{split}
&y_\phi \approx 3.6 \times 10^{-24} {m_{20}^{-{1 \over4}}\ \over N_c} {m_f \over m_e}
~~  \mbox{with}~~ m_{20} \ll{ 31\over N_c^{2\over5}} \left( {m_f \over m_e}\right)^{8\over5}
   ~~\mbox{for}~~  x_f \ll x_1 \ll 1 , \\
&  y_\phi \approx 5.3 \times 10^{-24} m_{20}  \left( {m_e \over m_f}\right)^{1\over2}
~~  \mbox{with}~~ m_{20} \gg 2\times 10^3   \left( {m_f \over m_e}\right)^{2}
~~\mbox{for}~~ 1\ll  x_f \ll x_1 .
\end{split}
\end{equation}
Let us remark that the first case in (\ref{y-m-II}) is the only one allowing the lightest dark matter mass in our mechanism, which could be constrained by various observations summarized in \cite{hayashi}.

For the second case of (\ref{y-m-II}), the BBN condition (\ref{bbn}) turns out to give stronger limit than (\ref{freezein}):
\begin{equation}
m_\phi \ll 10^{-6}\, (2.5\times 10^{-3})
 \mbox{ eV}~~\mbox{for}~~ f=e~(b),
\end{equation}
equivalently $y_\phi \ll 5.7 \times 10^{-10}\, (1.5\times10^{-8})$.

\medskip
\underline{Case III: $x_{ew} <  x_1 < x_f$ }
\medskip

This is a bit involved situation where there are three asymptotic regimes with the transition from $m_T^2 > m_\phi^2$ to $m_\phi^2 < m_T^2$ when the fermions are in thermal equilibrium. That is,  the evolution of the bosonic field is described by
\begin{equation}
\begin{split}
 \tilde \phi''(x) +  {3 \over 2 x}\tilde\phi'(x) + {x_1\over x} \tilde\phi(x)  \approx {x_S\over x}
 & ~~\mbox{for}~~ x_{ew}\ll x\ll x_1  \\
 \tilde \phi''(x) +  {3 \over 2 x}\tilde\phi'(x) +  \tilde\phi(x)  \approx {x_S\over x}
& ~~\mbox{for}~~ x_1\ll x\ll x_f \\
 \tilde \phi''(x) +  {3 \over 2 x}\tilde\phi'(x) + \tilde\phi(x)  \approx 0
& ~~\mbox{for}~~  x_f\ll x .
\end{split}
\end{equation}
Applying the previous results, one can find the corresponding solution in each regime which leads to the final oscillation solution at $x_{eq}$.  Its integration constants $C_{1,2}$ are complicated functions of $x_1$ and $x_F$. We just report the asymptotic values of $c_f^2$ in  the three different limits:
\begin{equation}
c_f^2 \approx \left\{
\begin{split}
 4 \sqrt{2}\,\Gamma\big({5\over4}\big)^2 x_f^2
 & ~~\mbox{for}~~ x_{1}\ll x_f \ll 1 \\
 {\pi \Gamma({3\over4})^2 \over 2\sqrt{2}}
& ~~\mbox{for}~~ x_1\ll 1\ll x_f \\
{\pi  \over3 x_1^{5/2}} +{\pi  \over2 x_f^{1/2}}
& ~~\mbox{for}~~  1\ll x_1 \ll x_f .
\end{split}\right.
\end{equation}
Thus, we obtain the following parameter ranges explaining the dark matter density.\\
(i) $ x_{1}\ll x_f \ll 1$:
\begin{equation}
 y_\phi \approx 3.6 \times 10^{-24} {m_{20}^{-{1\over4}} \over N_c}  {m_f \over m_e}
~~\mbox{with}~~ {28 \over N_c^{2\over5}} \left( m_f \over m_e\right)^{8\over5} \ll m_{20} \ll
\mbox{Min}\!\left[ 2\times10^3 \left( m_f \over m_e\right)^2, ~
{6\times10^5 \over N_c^{2\over5}} \left( m_f \over m_e\right)^{4\over5}\right]
\end{equation}
(ii) $ x_1\ll 1\ll x_f$:
\begin{equation}
y_\phi \approx 3 \times 10^{-27} { m_{20}^{3\over 4} \over N_c} {m_e \over m_f}
 ~~\mbox{with}~~ 2\times10^3\left( m_f \over m_e\right)^2 \ll m_{20} \ll
 \mbox{Min}\!\left[ 3.2\times 10^{11} N_c^2 \left( m_f \over m_e\right)^4, ~
 {2.1\times10^{16} \over N_c^{2}} \left( m_e \over m_f\right)^{4}\right]
\end{equation}
(iii) $1\ll x_1 \ll x_f$:
\begin{equation} \label{ymIIIc}
\begin{split}
&y_\phi \approx 1.9 \times 10^{-22} {m_{20}^{1\over3} \over N_c^{1\over6}} \left(m_f\over m_e\right)^{2\over3}
~\mbox{with}~{ 1.2\times10^8 N_c^{{5\over4}}}    \left( m_f \over m_e\right)^{13\over4}\ll m_{20} \ll 4.4\times 10^{11}  N_c^2\left( m_f \over m_e\right)^4
~\mbox{for}~  x_1^5 \ll x_f \\
&y_\phi \approx 4.5 \times 10^{-28} {m_{20} \over N_c} \left(m_e\over m_f\right)^{3\over2}
~~\mbox{with}~~  m_{20} \gg 4.9\times10^8 N_c^{5\over4}\left( m_f \over m_e\right)^{13\over 4}
~\mbox{for}~ x_f\ll x_1^5.
\end{split}
\end{equation}
In the first case of (\ref{ymIIIc}), the condition (\ref{freezein}) puts the stronger bound for $f=b$, e.g.,
$ m_{20} \ll 1.9\times 10^{22}$ and thus $y_\phi \ll 2.5\times10^{-9}$.
In the second case, additional upper bounds can be obtained from (\ref{bbn}) or (\ref{freezein}):
\begin{equation}
m_\phi \ll 1.3 \times 10^{-2}\, ( 6.1\times 10^3) \mbox{ eV} ~~\mbox{for}~~ f=e\,(b),
\end{equation}
requiring $y_\phi \ll 5.7 \, (1.2) \times 10^{-10}$. 

Let us finally remark that all the solutions obtained in this section satisfies the condition $ y_\phi^2 \phi^2 \ll T^2$ guaranteeing the starting equation for the  thermal effect (3). 

\section{Coherent oscillations from neutrinos}

Neutrinos remain relativistic and keep the thermal distribution even after the matter-radiation equality.
Thus, the bosonic oscillation evolves following (\ref{eq:phi}) down to $x_{eq}$. We find no parameter space explaining the proper dark matter abundance for $x_1 \ll x_{ew}$. In the opposite case, the evolution equation is given by 
\begin{equation}
\begin{split}
 \tilde \phi''(x) +  {3 \over 2 x}\tilde\phi'(x) + {x_1\over x} \tilde\phi(x)  \approx {x_S\over x}
& ~~\mbox{for}~~ x_{ew}\ll x \ll x_1\\
 \tilde \phi''(x) +  {3 \over 2 x}\tilde\phi'(x) +  \tilde\phi(x)  \approx {x_S\over x}
& ~~\mbox{for}~~ x_1 \ll x
\end{split}
\end{equation}
which is solved by
\begin{equation}
 \tilde\phi(x) =\left\{
 \begin{split}
&  {x_S\over x_1} \left[ 1-\sqrt{x_{ew}\over x} \cos\left(2(\sqrt{x x_1}-\sqrt{x_{ew}x_1})\right)
- {1\over 2 \sqrt{x x_1} } \sin\left( 2(\sqrt{x x_1}-\sqrt{x_{ew}x_1})\right)  \right]
~~\mbox{for}~~ x_{ew}\ll x \ll x_1\\
 &  C_1 {J_{1\over4}(x)\over x^{1/4} }   + C_2 {Y_{1\over4}(x) \over x^{1/4} }
 + x_S \left[ {\pi   \over \Gamma({3\over4})}  {J_{1\over4}(x) \over (2x)^{1/4} }\, G_1(x) -
{ \Gamma({3\over4}) \over 12 }  {J_{-{1\over4}}(x)  \over (2x)^{1/4} }\, G_2(x)  \right]
~~\mbox{for}~~ x_1 \ll x .
\end{split} \right.
\end{equation}
Thus we find
\begin{equation}
c_\nu^2 =
{\pi \Gamma({3\over4})^2  \over 2\sqrt{2}}+
{ 2^{3\over4} \pi^{3\over2} C_1 \over  \Gamma({1\over4})  x_S} +
{ C_1^2 +C_2^2\over x_S^2 } .
\end{equation}
Matching the boundary condition at $x_1$, the integration constants $C_{1,2}$ are given by
\begin{equation}
\begin{split}
&{C_1 \over x_S}=-{\pi \over 2^{1\over4} \Gamma({3\over4})} G_1(x_1)
                          +{2^{5\over4} \Gamma({3\over4}) \over 3} G_2(x_1)
-{\pi\over2 } {Y_{1\over4}(x_1)\sin^2(x_1)-Y_{-{3\over4}}(x_1)[x_1 -{1\over2}\sin(2x_1)] \over x_1^{3/4}} \\
& {C_2\over x_S}=-{2^{5\over4} \Gamma({3\over4}) \over 6} G_2(x_1)
+{\pi\over2 } { J_{1\over4}(x_1)\sin^2(x_1)-J_{-{3\over4}} (x_1) [x_1 -{1\over2}\sin(2x_1)]  \over x^{3/4}}
\end{split}
\end{equation}
and thus two asymptotic values of
$c_\nu^2= 3.34\, (1.67)$ for $x_1\gg 1\, (x_1\ll 1)$.
Now one can see that the right dark matter abundance is obtained for $x_1 \gg 1$ under the condition of
\begin{equation}
y_\phi \approx 4.2 \times 10^{-20} m_{20}^{3/4} \left( 0.05 \mbox{eV} \over m_\nu \right)
~~\mbox{with}~~ m_{20} \gg 710 \left( 0.05 \mbox{eV} \over m_\nu \right)^4.
\end{equation}
Let us remark that the lower bound on $m_\phi$ comes from the condition that the medium-induced mass-squared to be smaller than $T^2$, equivalently $(y_\phi \phi)^2 \ll T^2$,  all the time so that the standard cosmology is maintained \cite{cck}.
This corresponds to $x_1 \ll 0.3 x_{eq}$, which realizes an interesting option of  very
late production of dark matter population.

In the case of $\phi$-$\nu$-$\nu$ coupling, 
the BBN bound is calculated to be $y_\phi \lesssim 7 \times 10^{-6}$ \cite{escudero} which implies 
\begin{equation}
m_\phi  \lesssim 0.092 \mbox{ eV} \left( m_\nu \over 0.05 \mbox{ eV}\right)^{4/3}
\end{equation}
and thus the decay $\phi \to \nu \nu$ is automatically forbidden.

\section{conclusion}

When a scalar boson $\phi$   is assumed to couple dominantly to a quark or a lepton,
the thermal effect, encoded in the thermal mass $m_T$ and the temperature-dependent source term, plays an important role in the classical evolution of the boson field describing the coherent state.  Solving the evolution equation analytically within the sudden change approximation in various asymptotic limits, it is demonstrated that the bosonic coherent state can develop properly to constitute the dark matter density.
While the source term determines the overall normalization of the amplitude of the coherent oscillation,
various different types of solutions appear depending on the hierarchies among the initial temperature $T_{ew}$ at which the electroweak phase transition occurs,   $T_1$ at which $m_\phi = m_T$, and
 $T_f \equiv m_f$ below which the thermal effect of the fermion $f$ disappears.

\medskip
When the boson couples to a charged fermion, some of the allowed parameter space overlap with
the freeze-in production regime or conflict with the BBN constraint. Excluding these regions, the largest possible value of $m_\phi$ can reach up to $\sim$ MeV (16) when the boson couples to the bottom quark (also to the tau lepton). Interestingly there appears only one case (21) which allows the ultralight mass range, $m_\phi \lesssim 10^{-19}$ eV, and could be probed by astrophysical and cosmological observations. We also note that
the Yukawa coupling in a wide range of $y_\phi \sim 10^{-28:-8}$ is required depending on the situation.

\medskip
The boson-neutrino-neutrino coupling has a unique feature that it realizes a very late genesis of dark matter, even at the temperature close to the matter-radiation equality $T_{eq}$.
In general, $T_1$ residing between $T_{ew}$ and $T_{eq}$ has to be much smaller than $\sqrt{m_\phi M_P}$ to generate the right dark matter abundance, which occurs in a wide range of $m_\phi \sim 10^{-17:-1}$ eV  with $y_\phi \sim 10^{-17:-5}$ for $m_\nu = 0.05$ eV.

\bigskip

{\bf Note added:} While completing the draft, there appeared Ref.~\cite{batell} which proposes the same idea and overlaps partially with our analyses.


\end{document}